\documentclass[aps,prd,a4paper,%nofootinbib,
preprintnumbers,twocolumn,floatfix,showpacs]{revtex4}
\usepackage{graphicx,graphics,epsfig,epic,eepic}    
\usepackage[usenames,dvipsnames]{color}
\usepackage{subfigure}
\usepackage{dcolumn}    
\usepackage{amsmath,amsfonts,amssymb}
\usepackage{mathrsfs}   	
\usepackage{bm}              
\usepackage{psfrag}
\usepackage{units}      	
\usepackage[colorlinks=true,linkcolor=blue,citecolor=blue]{hyperref}
\usepackage{enumerate}
\usepackage{braket}
\newcommand{\1}{^\mathrm{in}}
\newcommand{\2}{^\mathrm{out}}
\newcommand{\ha}{\hat{a}}
\newcommand{\hb}{\hat{b}}
\newcommand{\hwk}{\hbar \omega_k}
\newcommand{\wkout}{\omega\2_k}
\newcommand{\La}{\lambda_{{k}}(t)}

\newcommand{\Ga}{\gamma_{{k}}(t)}

\newcommand{\abs}[1]{\vert #1 \vert}
\newcommand{\brv}[1]{\langle #1 \rangle}

\definecolor{Gruen}{rgb}{.322,.537,.035}

\DeclareMathOperator{\tr}{tr}

\begin{document}
%%%%%%%%%%%%%%%%%%%%%%%%%%%%%%%%%%%%%%%%%%%%%%
%%%%%%%%%%%%%%%%%%%%%%%%%%%%%%%%%%%%%%%%%%%%%%
% \preprint{gr-qc/xxx}
%%%%%%%%%%%%%%%%%%%%%%%%%%%%%%%%%%%%%%%%%%%%%%
%\title{Nonclassicality criteria for excitations in Bose-Einstein condensates?}
\title{On the observation of nonclassical excitations in Bose-Einstein condensates}
%%%%%%%%%%%%%%%%%%%%%%%%%%%%%%%%%%%%%%%%%%%%%%
\author{Andreas Finke$^{1,2}$, Piyush Jain$^{3}$ and Silke Weinfurtner$^{1,4}$}
\affiliation{
$^1$ School of Mathematical Sciences, University of Nottingham, University Park, NG7 2RD Nottingham, UK \\
$^2$ School of Physics \& Astronomy, University of Nottingham, University Park, NG7 2RD Nottingham, UK \\
$^3$ University of Alberta, Edmonton, Alberta, Canada T6G 2J1 \\
$^4$ Perimeter Institute for Theoretical Physics, Waterloo, Ontario, Canada N2L 2Y5}
%%%%%%%%%%%%%%%%%%%%%%%%%%%%%%%%%%%%%%%%%%%%%%

\begin{abstract} 
In the recent experimental and theoretical literature well-established nonclassicality criteria from the field of quantum optics have been directly applied to the case of excitations in matter-waves. Among these are violations of Cauchy-Schwarz inequalities, Glauber-Sudarshan P-nonclassicality, sub-Poissonian number-difference squeezing (also known as the two-mode variance) and the criterion of nonseparability. We review the strong connection of these criteria and their meaning in quantum optics, and point out differences in the interpretation between light and matter waves. We then calculate observables for a homogenous Bose-Einstein condensate undergoing an arbitrary modulation in the interaction parameter at finite initial temperature, within both the quantum theory as well as a classical reference.  We conclude that to date in experiments relevant for analogue gravity, nonclassical effects have not conclusively been observed and conjecture that additional, noncommuting, observables have to be measured to this end.
\end{abstract}

%%%%%%%%%%%%%%%%%%%%%%%%%%%%%%%%%%%%%%%%%%%%%%
% PACS numbers:
%%%%%%%%%%%%%%%%%%%%%%%%%%%%%%%%%%%%%%%%%%%%%%
\maketitle
%%%%%%%%%%%%%%%%%%%%%%%%%%%%%%%%%%%%%%%%%%%%%%
%
%%%%%%%%%%%%%%%%%%%%%%%%%%%%%%%%%%%%%%%%%%%%%%

%%%%%%%%%%%%%%%%%%%%%%%%%%%%%%%%%%%%%%%%%%%%%%
Bose-Einstein condensation (BEC) is a macroscopic quantum phenomenon where a large fraction 
of the bosons occupy the same lowest quantum state, and thus form a coherent matter wave. 
Many properties of the condensate can be captured by a Schr\"odinger-type equation for a complex field with a nonlinear potential referred to as the Gross-Pitaevskii equation (GPE)~\cite{Davis:1995qf,Castin2001,Polkovnikov:2009ys}. 
To cite one example, the GPE correctly predicts discrete values for the circulation of the velocity field in a BEC~\cite{Fetter:2009aa}. Nevertheless, within this description the evolution of the condensate can be considered a classical process~\cite{Johnson:2014aa}. On the other hand it is possible to excite small fluctuations of the condensate, for example by means of rapid changes in the condensate parameters~\cite{Bauer:2009fk,Inouye:1998kx,Jaskula2012,Steinhauer:2014aa,Steinhauer:2015aa,Steffens:2014zba}, which may activate their nonclassical behaviour. 
An ongoing line of research is to investigate experimentally the manipulation and detection of fluctuations in BECs, e.g. exploring their quantum nature~\cite{Krachmalnicoff2010,Jaskula2010,Kheruntsyan2012} and correlations~\cite{Schweigler:2015maa,Steffens:2014zba}.
This is partially motivated by various analogue gravity studies~\cite{Unruh:1981bi,Barcelo2011,Schutzhold:2007aa,Faccio:2013yq}, where BECs are utilized as quantum simulators~\cite{Johnson:2014aa} of quantum field theories in curved spacetime (QFTCS). In QFTCS one is interested in how quantum fields propagate on a classical curved spacetime geometry acting as a non-trivial background configuration. Within analogue gravity studies the ultimate goal is to mimic and capture genuine quantum effects predicted from QFTCS. 
The field of experimental analogue gravity in BECs started only a few years ago, and with recent advances, e.g.~mimicking black hole evaporation~\cite{Steinhauer:2014aa,Steinhauer:2015aa,Lahav:2009wx,Coutant:2010aa,Finazzi:2010nc} and cosmological particle production in our universe~\cite{Jaskula2010,Hung:2013aa,Jain2007,Carusotto:2010dz,Busch:2014ac}, and the debate on the nonclassicality of the observed effects is as timely as ever. Hence, it is essential to find suitable observables that can distinguish between classical and nonclassical features of excitations in the condensate. Although in principle possible, we are lacking a Bell-type experiment to rule out \emph{any} classical model (assuming local realism) for the excitations in a BEC, and in this sense establish their nonclassicality once and for all~\cite{Clauser:1974aa,Brunner:2014aa}. We partially address this issue, by investigating a more restrictive notion of nonclassicality. A system shows nonclassical behavior precisely when the \emph{observations} made are incompatible with the predictions of a \emph{specific} classical reference: a specified classical theory and the accessible observables.

Our approach is mainly motivated by the quantum optics revolution~\cite{Grynberg:2010aa}. Maxwell's theory of electromagnetism supplies us with an excellent classical description of light as a wave phenomenon. There are however situations where the quantum nature of light cannot be neglected. For example, antibunching experiments provide direct evidence for the existence of photons~\cite{Kimble1977,Paul:1982aa} (we explain this effect in more detail in the next section). Within quantum optics nonclassicality criteria have been defined that delineate between classical and nonclassical attributes of the quantum states. Since the quantum theories of both light and BECs can be described in a Bosonic Fock space, one can in principle apply nonclassicality criteria taken or inspired from quantum optics to the BECs. However, as we will demonstrate, one can not always carry over the interpretation as well. Below we demonstrate the pitfalls involved in even establishing nonclassicality with currently accessible observables in the sense of ruling out a \emph{given} classical theory.

We focus on the case of the finite-temperature homogeneous BEC undergoing an arbitrary parametric excitation due to a variation of the scattering length. From an experimental perspective this can be implemented straightforwardly, see for example~\cite{Bauer:2009fk,Inouye:1998kx,Hung:2013aa}. For the quantum description we assume quadratic Bogoliubov theory which neglects phonon interactions so that the non-equilibrium dynamics can be solved for exactly. It preserves the Gaussianity of the quantum state. As a classical reference we employ a quadratic semiclassical approach based on the GPE with an initial thermal mode population which also preserves Gaussianity. The quantum mechanical predictions for the moments of mode occupations approach the predictions of this classical theory in the high temperature limit.  \\

Motivated from quantum optics studies three kinds of inequalities were utilized in recent BEC experiments, with the goal to determine nonclassicality in atom optics experiments, namely sub-Poissonian statistics~\cite{Kheruntsyan:2002aa,Ogren:2009aa,Jaskula2010,Jaskula2012,Wasak:2014aa,Frank:2015aa}, and intensity~\cite{Olsen:2002aa,Marino:2008aa,Buchmann2010,Kheruntsyan2012,Wasak:2014aa,Nova:2014aa,Nova:2015ab,Nova:2015aa,Boiron:2015aa} and mode~\cite{Campo:2005aa,Busch:2013aa,Busch:2014ab,Finazzi2014,Busch:2014ac,Steinhauer:2015aa,Steinhauer:2014aa,Steinhauer:2015aa} Cauchy-Schwarz inequalities (CSIs). Among these three inequalities sub-Poissonian statistics is the most accessible variable, as it can be obtained directly from the normalised number difference variance, also referred to as the two-mode variance (TMV), see eqn.~(\ref{Eq:TMV_fun}-\ref{Eq:TMVquantum}). The intensity CSI involves \emph{normally} ordered density-density correlations of two symmetrically occupied modes, and is violated when the normally ordered cross-correlations exceed the normally ordered auto-correlations, see eq.~(\ref{Eq:intensityCSI}). In~\cite{Kheruntsyan2012} it was argued that the normally ordered density-density correlations are accessible in time-resolved time-of-flight (TOF) measurements. The mode Cauchy-Schwarz inequality, see eq.~(\ref{Eq:modeCSI}), is a comparison between \emph{anomalous} and mode density.   A direct measurement of the anomalous density is not possible, but recently an indirect measurement of it has been suggested~\cite{Steinhauer:2015ab,Steinhauer:2015aa}. 

As we shall see, from all of the above mentioned criteria \emph{only} the mode CSI is \emph{in principle} a sufficient measure of nonclassicality in the sense of ruling out our classical reference.
We also show that given some stronger assumptions, i.e.~the validity of the approximate quantum theoretical model (within the Bogoliubov approximation the quantum states for the excitations are Gaussian), all of the three criteria are mathematically equivalent, and since the mode Cauchy-Schwarz inequality is violated if and only if the state is nonseperable~\cite{Busch:2014aa}, all three of them can equally well facilitate as entanglement criteria. Nevertheless we show explicitly, when it comes to the experimental realisation of the
 above mentioned criteria, \emph{none} of them are sufficient measures of nonclassicality in the experimental setup under consideration here.

%%%%%%%%%%%%%%%%%%%%%%%%%%%%%%%%%%%%%%%%%%%%%%%%%%%%%%%%%%%%%%%%%%%
%
\section{Nonclassicality criteria to falsify a specific classical theory~\label{Sec:NonClassicalityCriteria}}
%
%%%%%%%%%%%%%%%%%%%%%%%%%%%%%%%%%%%%%%%%%%%%%%%%%%%%%%%%%%%%%%%%%%%
In quantum optics an important class of observables is provided by absorbing photodetection. 
As a case in point we consider monochromatic light emitted from a point source and photo detectors with no dead time that detect light of a specific wave vector. % and frequency.  
In the semiclassical theory of atom-light interactions~\cite{Bohr1924:1, Bohr1924:2} the joint click rates of a single or several photodetectors 
are directly proportional to the moments of mode intensities.  
In the theory of the quantized field the intensity $I_a$ of a mode $a$ is represented by the photon number operator $\hat n= \hat a^\dag \hat a$, and the joint click rates are proportional to \textit{normally and time ordered} moments of this operator~\cite{VogelWelsch}; 
e.g.~for $m$ coincident measurements at equal time and position they are proportional to $\braket{(\hat a^\dag)^m \hat a^m }= \braket{\hat n( \hat n-1) \hdots (\hat n-m+1)}$. The destructive measurement of photons results in decreased auto-correlations, while the classical reference assumes a ``non-destructive'' measurement of waves resulting in $\langle n^m \rangle_{cl}$, where $n=I$ is the (fluctuating) mode intensity. 
 For example consider the observation of intensity correlations $C(\tau)= \mathbb{E}( I(t)I(t+\tau) )$ 
 of continuously emitted light (i.e.~in a stochastic steady-state) from a fluorescing atom at two subsequent times separated by $\tau$ as recorded by a single detector. 
(We use the notation $\mathbb{E}(...)$ for mean values of observables as measured in the laboratory.)
If the density-density correlations are increasing with $\tau \ge 0$, we can rule out the classical model because the Cauchy-Schwarz theorem would always imply $dC/d\tau(0) \le 0$~\cite{VogelWelsch}. However, the observation of $dC/d\tau(0)>0$, known as \emph{photon antibunching},
 was reported in~\cite{Kimble1977}. This effect is genuinely nonclassical by definition because it is not understood within the classical reference. Within quantum theory, the observed data can be explained by the fact that single photons are arriving with a tendency of
 being separated from each other and the normally ordered correlator is strongly decreased. The criterion for
 quantum states allowing for the observation 
 of $dC/d\tau(0)>0$ can be formulated as the inequality
\begin{equation} \label{Eq:CSI-Single-Mode}
\langle \hat a^\dag \hat a^\dag \hat a \hat a \rangle < \langle \hat a^\dag \hat b^\dag  \hat b \hat a \rangle,
\end{equation} 
where $\hat a=\hat a(t)$, $\hat b=\hat a(t+\tau)$. It follows that this expression
serves as a sufficient \emph{nonclassicality criterion} for the quantum state in an experiment of the type outlined above. 

The concept of nonclassical states and any corresponding criteria can be extended 
to the case of multiple detectors. 
In particular, let  $\ket{\alpha, \beta}$ be a tensor product of coherent states for two modes $a, b$ 
entering two detectors at different positions, and consider 
the measurement of coincident click rates, proportional to normally ordered correlators of $\hat n_a, \hat n_b$ in the quantum theory.
Further, let us call states  $\rho = \int d\alpha d\beta P(\alpha,\beta) \ket{\alpha,\beta} \bra{\alpha,\beta}$
\emph{P-classical} if their Glauber-Sudarshan P-representation $P(\alpha, \beta)$ 
has the properties of a classical (i.e.~positive) probability distribution for the two complex mode amplitudes. 
Using the defining property that the coherent states are eigenstates of their corresponding mode annihilation operators,
it is evident that one can reproduce any coincident click rate within the classical reference by using $P(\alpha,\beta)$ as the probability distribution 
for the complex classical mode amplitudes~\cite{Titulaer1965, Mandel1986, VogelWelsch}. On the other hand, 
for \emph{P-nonclassical} states without such a P-representation there should exist in principle 
a coincidence counting experiment falsifying the classical theory. Thus, in the case of measuring such counting rates in quantum optics, a state is classical if and only if it is P-classical. 
A sufficient criterion to establish P-nonclassicality of a state is similarly given by inequality~(\ref{Eq:CSI-Single-Mode}), 
where now $a$ and $b$ are two different modes at equal time, as is proven by contradiction from the Cauchy-Schwarz theorem.
This inequality is equivalent to 
\begin{equation} \label{Eq:intensityCSI} 
G^{(2,2)}_{a,b}>(G^{(2,2)}_{a,a}G^{(2,2)}_{b,b})^{1/2}, 
\end{equation}
for the special case that $G^{(2,2)}_{a,a} = G^{(2,2)}_{b,b}$. Here 
$G^{(2,2)}_{{a}{b}} \equiv \brv{\ha^\dag \hb^\dag \ha \hb} $
are the normally ordered correlators. 
We therefor refer to this criterion as the violation of the \emph{intensity CSI}.

 We now discuss the connection of the intensity CSIs and another correlation measure, the two-mode variance (TMV),
 also called the number squeezing parameter~\cite{Ogren2009, Buchmann2010}, which is experimentally accessible by repeatedly measuring the intensity of the two modes. 
The TMV is defined as the normalized number difference variance
\begin{equation} V=Var(I_a-I_b)/\mathbb{E}(I_a+I_b) \ge 0
\end{equation}
for two modes labelled $a$ and $b$. 
Similar to referring to the case of a single variable 
with a variance smaller than its mean as sub-Poissonian, we 
 refer to $V<1$ as sub-Poissonian 
statistics in accordance with the literature. 
However $V$ is sensitive to both the single-mode statistics 
and two-mode correlations. 
While $V=1$ for the state $\ket{\alpha, \beta}$, in general neither uncorrelated nor Poissonian 
 variables are necessary for $V=1$ \footnote{An example are the binomially distributed, 
 anti-correlated numbers of particles found in two non-overlapping regions in a box filled with an ideal gas.
 Simple examples with (positive) correlations also exist.}. 
 For the symmetric case, where $\mathbb{E}((I_a)^n)=\mathbb{E}((I_b)^n)$ for $n=1,2$, 
 the TMV simplifies to 
 \begin{equation} 
 \label{Eq:TMV_fun}
  V=\frac{\mathbb{E}(I_aI_a)-\mathbb{E}(I_aI_b)}{\mathbb{E}(I_a)}\,.
  \end{equation}
 In a \emph{quantum theory} this can be written as  
\begin{eqnarray} \label{Eq:TMVquantum}
 V= 1+\frac{G_{{a},{a}}^{(2,2)} - \,G_{{a},{b}}^{(2,2)}}{\brv{\hat{n}_{a}}}  
 \label{Eq:V2},
 \end{eqnarray}
so that it follows 
that $V<1$ and a violation of the intensity CSI are equivalent, as was noted previously in~\cite{Kheruntsyan2012}. 
Note that since $V\ge0$, equation~(\ref{Eq:TMV_fun}) also proves that cross-correlations never
 exceed the \emph{non-normally} ordered auto-correlations; the normal ordering is crucial for a violation of the intensity CSI. 
In a quantum theory, due to the non-zero commutator, the normal ordering reduces the density-density auto-correlations, such that $G_{{a},{a}}^{(2,2)}=\brv{\hat{n}_{a}^2}-\brv{\hat{n}_{a}}$, which allows for $G_{{a},{a}}^{(2,2)} < G_{{a},{b}}^{(2,2)}$.
 
For the BEC, the implication that P-nonclassical states 
can demonstrate nonclassical behavior is not generally true, 
as it is fundamentally based on the ability to measure normally 
ordered moments of intensity in the quantum theory 
while at the same time these are represented without normal order in the classical reference. 
We will see below that this is not the case in the BEC. Thus the criteria for nonclassical states 
from quantum optics based on establishing P-nonclassicality, 
such as the violation of the intensity CSI~(\ref{Eq:intensityCSI}), 
are neither \emph{a priori} related to nonclassicality.

To further explore these issues we focus on the specific example of (parametrically) excited, strongly correlated two-mode number fluctuations in a BEC~\cite{Jaskula2012}. To determine the validity of the nonclassicality criteria for this system it is necessary to compare both the quantum and the corresponding semiclassical theories, as described in Secs.~\ref{Sec:QuantumTheory} and \ref{Sec:ClassicalTheory} respectively.
 
%%%%%%%%%%%%%%%%%%%%%%%%%%%%%%%%%%%%%%%%%%%%%%%%%%%%%%%%%%%%%%%%%%%
%
\section{Quantum theory \label{Sec:QuantumTheory}} 
%
%%%%%%%%%%%%%%%%%%%%%%%%%%%%%%%%%%%%%%%%%%%%%%%%%%%%%%%%%%%%%%%%%%%
We consider an interacting Bose gas in a box of volume $V = L^3$ 
with time-dependent interaction strength. 
The second quantized Hamiltonian is given by 
\begin{eqnarray}
\label{Eq:Hfourier}
\hat{H}(t) = \sum_{\mathbf k} \frac{\hbar^2 k^2}{2 m} \hat{a}_{\mathbf k}^\dagger \hat{a}_{\mathbf k} + \frac{U(t)}{2 V} \sum_{{\mathbf k}{\mathbf k}'{\mathbf q}} \hat{a}_{{\mathbf k}+{\mathbf q}}^\dagger \hat{a}_{{\mathbf k'}-{\mathbf q}}^\dagger \hat{a}_{\mathbf k} \hat{a}_{\mathbf k'} \,,
\end{eqnarray}
where $\hat{a}_{\mathbf k}$ annihilates a single particle eigenstate of the (translationally invariant) momentum operator and
 $\mathbf k \in 2 \pi\mathbb{N}^d/L$ .
Here we assume the box to be large enough
to approximate its single particle ground state by $\hat{a}_{\mathbf 0}^\dagger \ket{0}$.
For atoms of mass $m$ and s-wave scattering valid in the ultracold regime, the effective interaction is $U(t) =  4\pi n_0 \hbar^2 a_s(t)/m $.
In experiments, the time-dependence of the scattering length $a_s(t)$ can be achieved by use of an appropriate Feshbach resonance~\cite{Inouye:1998kx,Roberts:2001uq}.
We assume all changes to be slow enough not to excite the bound state of the resonance~\cite{Corson2015bound}.  

%%%%%%%%%%%%%%%%%%%%%%%%%%%%%%%%%%%%%%%%%%%%%%%%%%%%%%%%%%%%%%%%%%%
\subsection{Quasiparticles\label{Sec:ClassicalTheory}} 
%%%%%%%%%%%%%%%%%%%%%%%%%%%%%%%%%%%%%%%%%%%%%%%%%%%%%%%%%%%%%%%%%%%
Our formulation is based on standard Bogoliubov theory valid for a weakly interacting Bose gas \cite{Dalfovo:1999cl}. The Hamiltonian can be diagonalized by retaining only interaction terms to quadratic order in the total number of particles $N$ after the  Bogoliubov approximation 
\begin{equation} \label{Eq:Bapprox}
\ha_{\mathbf 0}  = \ha_{{\mathbf 0}}^\dag = \sqrt{N-\Delta}. 
\end{equation}
With the critical assumption of a small depletion of the condensate,
i.e.~$ \Delta  = \sum_{k\neq0} \langle \hat{a}_{\mathbf k}^\dagger \hat{a}_{\mathbf k} \rangle \ll N$, the Hamiltonian becomes a quadratic expression in the mode operators. Only modes with opposite momenta are coupled in a homogenous condensate and we henceforth use the abbreviated notation $\hat{a} \equiv \ha_{\mathbf k}$ and $\hat{b} \equiv \hat{a}_{-\mathbf k} $. A two-mode squeezing represented by the transformation 
\begin{equation}
\begin{aligned} 
\label{Eq:bff1}
a      \;\; &=& &u_k(t) \, \hat{A}&  \! \! \! + \;\; &v_k(t) \, \hat{B}^\dag  \,; \\
b^\dag  \;\; &=& &u_k(t) \, \hat{B}^\dag& \! \! \! + \;\; &v_k(t) \, \hat{A} \, ,
\end{aligned} 
\end{equation}
can be used to diagonalise the Hamiltonian. The coefficients may be chosen as real numbers satisfying $u_k^2-v_k^2=1$. Then the transformation is a Bogoliubov transformation, i.e.~leaves 
the bosonic commutation relation unchanged,  $[\hat{A}, \hat{A}^\dag] = 1 = [\hat{B}, \hat{B}^\dag]$, and $[\hat{A},\hat{B}] = 0$. 
Neglecting the ground state energy the Hamiltonian $\hat{H}(t) = \sum_{\mathbf{k}} \hwk(t) \hat{A}^\dag (t)\hat{A}(t)$ counts the elementary excitations
of the system, the so-called quasiparticles, with
\begin{eqnarray}
u_k(t), v_k(t) 
\label{Eq:ukvk}
 &=&  \frac{1}{2} \left( \frac{e_k^\mathrm{kin}}{e_k(t)} \right)^{1/2} \pm \frac{1}{2} \left( \frac{e_k(t)}{e_k^\mathrm{kin}} \right)^{1/2} \,,
\end{eqnarray} 
with density $n=N/V$,  kinetic energy $e_k^\mathrm{kin} = \hbar^2 k^2/2m$ and $e_k(t) = \hwk(t) = \sqrt{e_k^\mathrm{kin}
 \left( e_k^\mathrm{kin} + 2 U(t) n  \right)}$, the well-known Bogoliubov dispersion relation. 
 Assuming the equilibrated isolated system can be treated in the canonical ensemble with respect to the particles (or grand canonical with respect to the quasiparticles with vanishing chemical potential
 as their number is not conserved), with the density operator
 $\hat\rho_{th} \propto \exp{(-\beta \hat{H})}$, 
\begin{equation}
\label{eq:nkth}
\langle \hat{A}^\dag \hat{A} \rangle_{th} = \frac{1}{\exp(\hbar \omega_k / k_B T)-1} =: n_k^{th},
\end{equation}
and the anomalous quasiparticle average $\langle \hat{A} \hat{B} \rangle_{th}=0$.
For continuous time dependence of $U(t)$ the Heisenberg equations for the operators are, with 
 $S(t)= (A(t),B(t)^\dag)^T$, given by $dS/dt  = M(t) S$
where
\begin{equation}
\label{eq:firstordersys}
M(t) :=   \begin{pmatrix}
&-i \omega_k  &  \frac{d \log \sqrt{\omega_k}}{dt}\\
&\frac{d \log \sqrt{\omega_k}}{dt} &i\omega_k  
\end{pmatrix} .
\end{equation}
This implies that it is possible in general to describe the time evolution by a further linear 
transformation of the form  
\begin{equation}
\begin{aligned}
\label{Eq:bff3}
\hat{A}(t\2) \;\; &=& &\alpha_k^* \, \hat{A}(t\1)& \!\!\! + \;\; &\beta_k \, \hat{B}^\dag(t\1)\; ;&  \\
\hat{B}^\dag(t\2) \;\; &=& &\alpha_k \, \hat{B}^\dag(t\1)& \!\!\! + \;\; &\beta_k^* \, \hat{A}(t\1) \; .&
\end{aligned}
\end{equation}
The complex coefficients $\alpha_k^*$, $\beta_k$ and $\alpha_k$, $\beta_k^*$ are the entries of the first and of the second row of the fundamental matrix $\varphi(t=t\2)$
 of system~(\ref{eq:firstordersys}) respectively, i.e.~$d\varphi(t)/dt = M(t) \varphi(t)$ with $\phi(t\1) = id$. 
 Using $\det(\varphi) = \exp(\tr(\log(\varphi)))$ we have $d/dt(|\alpha_k|^2-|\beta_k|^2) = d/dt \det(\varphi) = \det(\varphi) \tr M(t) = 0$ 
 so that $|\alpha_k|^2-|\beta_k|^2 = 1$ at all times and (\ref{Eq:bff3}) is a Bogoliubov transformation. 
 We choose with no loss of generality $t\2=0$ such that for $t>0$ the interaction strength is kept constant, defining the \textit{out}-region, 
 in which the time evolution is a trivial phase oscillation. 
Note that the arbitrary time dependence of $U(t<0)$ is 
 now encoded in the complex Bogoliubov coefficients $\alpha_k$ and $\beta_k$ subject to the normalisation constraint.
  The dependence of these coefficients on $U(t<0)$ is implicit in our notation. 
 
By a periodic modulation of the interaction strength, $|\beta_k|$ grows exponentially with the number of periods if $k$ is in a region of instability. For example, for low amplitude sinusoidal modulations of the form $U(t) = U_0 (1+A \sin(\omega_D t))$ with $A\ll1$ the first resonance occurs at $\omega_k = \omega_D/2$~\cite{Staliunas2002}. More generally, some algebra shows the unstable regions are given for $\operatorname{\mathbb{R}e}\{\alpha_k^{(1)}\} > 1$, where  $\alpha_k^{(1)}$ is taken from the solution of system~(\ref{eq:firstordersys}) integrated for a single period.  Analytic results may be obtained in the case of a square wave modulation, for which we denote the amplitude $\pi A/4$. Furthermore, the position and growth rate of the first resonance~\footnote{Approximately given for $k$ solution of $\omega_k(U_{max}) + \omega_k(U_{min}) = \omega_D$ 
 such that $U_{max/min} = U_0(1\pm\pi A/3)$ for $U(t) = U_0 (1+A \sin(\omega_D t))$, which can be solved analytically.} of a sinusoidal modulation with a large amplitude can be approximated by a series of sudden changes of this kind. 
 Interestingly, this also predicts the positions of 
 extremely fast growing parametric resonances occurring only
 for certain, large enough $A$. Fast growing, nonperturbative parametric resonances 
 have been suggested to cause preheating in the reheating process of the inflationary universe~\cite{Kofman1994}.
 Note however that such resonances increase the condensate depletion dramatically and will eventually 
 lead to violation of the assumed linear theory. 
  Parametric resonance for the nonlinear, classical problem has been studied analytically for a
 variation of the trap in~\cite{Garcia1999} and numerically for a variation of the scattering length 
 in~\cite{Cairncross2014}.
 For the purpose of the present work we may consider a narrow resonance due to sinusoidal modulation with $A \ll 1$. Even when the first resonance mode is strongly excited, the particle number can be insignificant with respect to the total depletion for a system with many modes.
 
%%%%%%%%%%%%%%%%%%%%%%%%%%%%%%%%%%%%%%%%%%%%%%%%%%%%%%%%%%%%%%%%%%%
\subsection{Real particles}
%%%%%%%%%%%%%%%%%%%%%%%%%%%%%%%%%%%%%%%%%%%%%%%%%%%%%%%%%%%%%%%%%%%
We are primarily concerned with the momentum-space observables, 
which are the expectation values of products of time-dependent real particle operators of the form $a(t>0)$ and/or its conjugate, 
as these observables can be measured in current experiments, e.g.~using the standard time-of-flight method~\cite{Anderson1995, Davis:1995qf}. Note, this is in contrast to most of the analogue gravity studies on similar subjects, which focus on the quasi-particles modes, e.g.~\cite{Barcelo2011,Steinhauer:2015ab,Busch:2013aa,Busch:2014ac,Busch:2014ab}.
Combining the interaction squeezing~(\ref{Eq:bff1}) with the parametric excitations for $t<0$~(\ref{Eq:bff3})
and the subsequent phase evolution, we have $\hat{a}=  \lambda_k(t)^* \, \hat{A}\1 + \gamma_k(t) \, \hat{B}\1 {}^\dag$ and 
$b^\dag =  \lambda_k(t) \, \hat{B}\1{}^\dag + \gamma_k^*(t) \, \hat{A}\1$.
Because the Bogoliubov transformations form a group,
 the coefficients of the total transformation satisfy $\abs{\La}^2-\abs{\Ga}^2=1$, and are given by
\begin{eqnarray}
\label{Eq:lambda}
\lambda_k(t) &=&   u_k \2 \alpha_k \mathrm{e}^{i \wkout t} + v_k \2 \beta_k \mathrm{e}^{-i \wkout t} ; \\ 
\label{Eq:gamma}
\gamma_k(t) &=&  u_k \2 \beta_k \mathrm{e}^{-i \wkout t} + v_k \2 \alpha_k \mathrm{e}^{i \wkout t} ,
\end{eqnarray} and a straightforward calculation shows that 
\begin{eqnarray}
\label{Eq:gammasq}
\abs{\Ga}^2&=&(v_k\2)^2+\abs{\beta_k}^2+2(v_k\2)^2 \abs{\beta_k}^2 \times  \nonumber \\
&\times&\left(1-\kappa_k \cos \left[2\wkout t+ \delta_k\right]\right), 
\end{eqnarray}
where $\kappa_k=[(1+(v_k\2)^{-2})(1+|\beta_k|^{-2})]^{\frac{1}{2}}$ and $\delta_k = \arg \alpha_k - \arg \beta_k$.

Starting from a thermal initial state, the number of particles in mode $\mathbf{k}$
for $t>0$ is given by
\begin{eqnarray} 
\label{Eq:Nk}
n_k(t)&:=&\brv{\hat{a}^\dag \hat{a} }  =  n_k^{th} + \abs{\Ga}^2 + 2  n_k^{th} \abs{\Ga}^2 .
\end{eqnarray}
Here $n_k^{th}$ refers to the initial population when $U=U(t\1)$. 
With expression~(\ref{Eq:gammasq}) the depletion
\begin{equation}  \label{Eq:depletion}
 \Delta  = \sum_{\mathbf k} n_k(t)
\end{equation}
is when time averaged, 
\begin{eqnarray} 
\overline{ \Delta }&=& \sum_{\mathbf k} \left[ n_k^{th} + v_k\2 {}^2 + |\beta_k|^2 \right. \nonumber \\ 
 &+& \left.  2 (n_k^{th} v_k\2{}^2 + n_k^{th} \beta_k^2 +v_k\2{}^2  \beta_k^2)  \right. \nonumber \\
&+& \left.   4 n_k^{th} v_k\2{}^2  \beta_k^2 \right],  \label{eq:Nkmean}
\end{eqnarray}
in particular containing the following three terms: the \emph{thermal depletion} $n_k^{th}$ which vanishes only for $T\1=0$, the \emph{depletion due to interactions} $v_k\2 {}^2$ which vanishes only for $U\2=0$, and the \emph{quasiparticle production} $\beta_k^2$ which vanishes only for $U=const$. 
Furthermore all three mutual dual products of these terms appear as well as the triple product, signifying mutual amplification of these processes. 
For large but finite volume it is possible to stay in the validity regime of the Bogoliubov theory $\Delta \ll N$ by 
reducing the scattering length sufficiently. Significant interactions are still possible by increasing the particle density. 
For the rest of this paper, we assume that we are in this regime. 

The only other non-vanishing correlation containing two operators is 
\begin{eqnarray} 
\label{Eq:Mk}
m_k(t) &:= &\brv{a b }  =  \gamma_k(t) \lambda_k^*(t) (1+2 n_k^{th}) .
\end{eqnarray}
The modulus squared  
\begin{equation}
|m_k(t)|^2   = |\gamma_k(t)|^2(1+|\gamma_k(t)|^2) (1+2 n_k^{th})^2 \, ,
\end{equation}
is referred to as the \emph{anomalous} density. As we shall see the anomalous density plays an important role in terms of establishing the nonseparability for the quantum fluctuations in a BEC.

%%%%%%%%%%%%%%%%%%%%%%%%%%%%%%%%%%%%%%%%%%%%%%%%%%%%%%%%%%%%%%%%%%
\subsection{Correlations} 
\label{Subsec:QCorr}
%%%%%%%%%%%%%%%%%%%%%%%%%%%%%%%%%%%%%%%%%%%%%%%%%%%%%%%%%%%%%%%%%%
The two-mode squeezed thermal state we are considering here is the exponential of a quadratic expression in the mode operators. 
For such Gaussian states, a finite temperature Wick theorem exists for mode operator moments 
in anti-normal, normal, and symmetrized order, respectively. 
Higher moments can be computed by simply summing over all possible pairings of second moments of the mode operators. 

This immediately gives the normal ordered correlation functions, 
where for clarity we reintroduce the $\mathbf k$-dependence, as 
 \begin{equation} 
 G_{{\mathbf k},{\mathbf k}'}^{(2,2)}(t) = \!
 \left\{ \!\!
 \begin{array}{l l l}
 2 n_k^2(t) 		&  {\mathbf k}'={\mathbf k}  \\
 n_k^2(t) + |m_k^2(t)|^2 &{\mathbf k}'=-{\mathbf k} \\ 
 n_k^2(t) 		 & \mbox{else} \\
 \end{array}
 \right. . \label{Eq:Gkk}
 \end{equation} 
In our abbreviated notation, $G^{(2,2)}_{{a}{a}} = 2 n_k^2(t)$ and $G^{(2,2)}_{{a}{b}} = n_k^2(t) + |m_k^2(t)|^2 $ .
Then, the two-mode variance~(\ref{Eq:TMVquantum}) is    
\begin{eqnarray} \label{Eq:V1}
V &=& 1 + \frac{n_k^2 - |m_k|^2}{n_k} .
\end{eqnarray} 
Thus, we have sub-Poissonian statistics and violation of the CSI if and only if $|m_k|^2>n_k^2$.
This is an entanglement criterion
which was first derived in the context of cosmological inflation~\cite{Campo:2005aa} and subsequently applied to the \emph{phonons} 
of quenched and parametrically excited BECs in~\cite{Busch:2013aa, Busch:2014ac}. 
The connection of entanglement to the violation of the CSI has been noticed previously~\cite{Busch:2014aa}.  
 We can rewrite expression~(\ref{Eq:V1}) as
 \begin{eqnarray} 
 \label{Eq:V2}
   V =\frac{n_k^{th}(1+n_k^{th})}{n_k^{th}(1+2|\gamma_k|^2) + |\gamma_k|^2},
 \end{eqnarray}
 and in particular simplify the condition $V<1$ to obtain
\begin{equation}
|\gamma_k^2(t)| > n_k^{th}(T/2) \label{Eq:entangled2}.
\end{equation} 

Taking the limit of vanishing interaction $U\2\rightarrow 0$ quasiparticles reduce to real particles and the term $|\gamma_k^2(t)| $ may be replaced by $|\beta_k^2|$ 
in~(\ref{Eq:entangled2}). Then this expression agrees with the \emph{phonon} entanglement criterion found in~\cite{Bruschi:2013fk}. 

We now give an independent (and short) derivation that~(\ref{Eq:entangled2}) 
is a nonseparability criterion for the real particles,~i.e. atoms, of opposite momenta.  

For a given partition of the Hilbert space into each of the two modes and the rest of the system $\mathcal H = \mathcal H_a \otimes \mathcal H_b \otimes \mathcal H_r$ 
we may ask if a product state $\rho = \rho_{ab} \otimes \rho_r$ is separable with respect to the first two spaces. 
This means by definition 
that the bipartite state can be 
written as a mixture of products (i.e.~a convex sum), $\rho_{ab} = \sum_i p_i \rho^i_a \otimes \rho^i_b$  
with $\sum_i p_i = 1$ and $p_i > 0$, and the subsystems' states $\rho^i_m$ might be taken to be pure without loss of generality.
If this is not possible one has a nonseparable or entangled state with respect to $  \mathcal H_a \otimes \mathcal H_b$~\cite{Werner:1989aa}.

A requirement for a separable state is that the partial transpose of the density operator is again positive semi-definite~\cite{Peres1996}.
 If this is not the case the state must be nonseparable.
For continuous variable two-mode systems, like for the two modes $a,b$ considered here, 
the implication of this sufficient nonseparability criterion for the covariance matrix is also necessary for nonseparability if the state is a Gaussian Wigner function~\cite{Simon:2000aa}.
 A straightforward procedure to evaluate this entanglement criterion conveniently written in terms of moments of the annihilation and creation operators 
 directly is to consider the determinant~\cite{Shchukin:2005aa} 
\begin{equation}
D_5= 
\begin{vmatrix}
1 & \langle \hat a\rangle & \langle \hat b^\dag\rangle & \langle \hat b\rangle  & \langle \hat a^\dag\rangle    \\
\langle \hat a^\dag\rangle  & \langle \hat a^\dag \hat a\rangle & \langle \hat a^\dag \hat b^\dag\rangle  & \langle \hat a^\dag \hat b\rangle  & \langle \hat a^{\dag 2}\rangle   \\
\langle \hat b\rangle  & \langle \hat a\hat b\rangle & \langle \hat b^\dag \hat b\rangle  & \langle \hat b^2\rangle  & \langle \hat a^\dag b\rangle    \\
\langle \hat b^\dag\rangle  & \langle \hat a\hat b^\dag\rangle  & \langle \hat b^{\dag 2}\rangle   & \langle \hat b \hat b^\dag\rangle  & {\langle \hat a^\dag \hat b^\dag \rangle}     \\
\langle \hat a\rangle  & \langle \hat a^2\rangle & \langle \hat a \hat b^\dag\rangle & \langle \hat a\hat b\rangle  & \langle \hat a \hat a^\dag\rangle    
\end{vmatrix} .
\end{equation}
Here rows and columns have been reordered such that $D_5$ trivially splits into the product of two sub-determinants for our case,
\begin{equation}
D_5 = (n_k^2-|m_k|^2)((n_k+1)^2-|m_k|^2).
\end{equation}
The state is entangled iff $D_5<0$~\cite{Shchukin:2005aa}. Since in general $|\langle a b\rangle|^2 < \langle n_a\rangle \langle n_b+1 \rangle$~\cite{Adamek2013} the second factor is always positive and we have simply 
$|m_k|^2>n_k^2 \Leftrightarrow \text{entanglement} $, completing the proof. 

Since for any mixed separable state $|\langle ab \rangle|^2 \le \langle a^\dag a \rangle \langle b^\dag b\rangle$,
$|m_k|^2>n_k^2$ 
implies nonseparability even for non-Gaussian states~\cite{Shchukin:2005aa,Hillery:2006aa}. Hence the nonseperability criterion is \emph{equivalent} to a mode Cauchy-Schwarz inequality violation,
\begin{equation} \label{Eq:modeCSI}
\vert \langle \hat{a}_{k} \hat{a}_{-k} \rangle \vert ^2 >    \langle \hat{a}_{k}^\dag \hat{a}_{k} \rangle      \langle \hat{a}_{-k}^\dag \hat{a}_{-k} \rangle  ,
\end{equation}
applicable to Gaussian and non-Gaussian states.
The intensity Cauchy-Schwarz violation~(\ref{Eq:intensityCSI}) 
itself on the other hand does not imply nonseparability in general when the moments do not factorize as for 
Gaussian states. For example, $V=0$ for (mixtures of) 
Fock states $\ket{n}\otimes\ket{n}$ with the same number of particles in mode $a$ and $b$.

We see from~(\ref{Eq:V2}) that for the thermal state of the non-interacting system with $\gamma_k = 0$ 
the fluctuations are always super-Poissonian, $V = 1 + n_k^{th}$, and thus the state is separable. Increasing $|\gamma_k|$
due to quantum depletion or parametric excitations however leads to a decrease of $V$ and sub-Poissonian statistics and nonseparability are possible. 

For the interacting system in equilibrium, i.e.~without parametric excitation, 
it can be shown that $V$ as a function of $k$ increases with decreasing wave number $k$. 
For $k\rightarrow 0$, $V=1$ is reached for $k_B T = U n$, i.e.~the temperature is of the order of the chemical potential $\mu$ of the Bose gas. \emph{Here the atoms with opposite momenta are nonseparable already for all $k$ due to interactions.} This effect is absent for the quasiparticles.  

The term $|\gamma_k|^2$ on the left hand side of formula~(\ref{Eq:entangled2}) appears in~(\ref{Eq:Nk}). Within~(\ref{Eq:Nk}) it may be interpreted as  
a quantum, `spontaneous emission' term, comprising both quasiparticle production and quantum depletion. Note that for high temperatures $n_k^{th}(T/2)\gg 1$
Bogoliubov theory implies a spontaneous emission term $|\gamma_k^2(t)| \gg 1$ and entanglement when sufficiently strong correlations $V<1$ are measured,
but it also predicts that the spontaneous emission is insignificant compared to the amplification of thermal noise.  
In the next section we show that the same measurable strong correlations $V<1$ which imply entanglement in Bogoliubov theory are possible in a classical theory without a notion of entanglement. In this theory the spontaneous emission term is absent and only amplification of the thermal occupation occurs.

%%%%%%%%%%%%%%%%%%%%%%%%%%%%%%%%%%%%%%%%%%%%%%%%%%%%%%%%%%%%%%%%%%%

\section{Classical theory\label{Sec:ClassicalTheory}} 
%
%%%%%%%%%%%%%%%%%%%%%%%%%%%%%%%%%%%%%%%%%%%%%%%%%%%%%%%%%%%%%%%%%%
When quantum and thermal fluctuations are neglected, the mean-field dynamics of a Bose-Einstein condensate are well-described by the so-called Gross-Pitaevskii equation (GPE), given by: 
\begin{equation}
i \hbar \partial_t \phi = \left(-\frac{\hbar^2 }{2m} \nabla^2 + V(r) + U(t) \vert \phi \vert^2  \right)  \phi, 
\end{equation}
where $\phi(t,x)=\langle \hat \Psi(t,x) \rangle$ is the complex-valued macroscopic wave function and $V(r)$ is the trapping potential taken to be zero for the homogeneous case considered here. 
Formally the GPE can be obtained from the Heisenberg equation of motion $i \partial_t \hat\Psi = [\hat\Psi, \hat H]$ 
in the quantum theory and by a replacement $\hat\Psi \rightarrow \phi $ in the operator equation of motion, 
as is well known. For a more detailed derivation see for example~\cite{Castin2001}.

A suitable classical Hamiltonian yielding the GPE from the equation of motion is given by the 
mean field energy functional~\cite{Pethick:2001aa}
\begin{equation}
H[\phi] = \frac{1}{2} \int d\mathbf r \left(\hbar^2 \nabla \phi^* \nabla \phi/m + U(t)\int d{\mathbf r}' |\phi|^2 |\phi|^2 \right)
\end{equation}
with the understanding that the complex field $\phi(x) \equiv (Q(x) - i P(x))/\sqrt(2)$.
Indeed with this definition Hamiltons equations $\partial_t Q =  \delta H/\delta  P $ and $\partial_t P =  - \delta H / \delta  Q $  appear in the form $i \partial_t \phi =  \delta H / \delta \phi^* = \{\phi, H\}$,
and the right hand side of this expression amounts to the same formal manipulations as in the right hand side of the 
 Heisenberg equation of motion $i \partial_t \hat\Psi = [\hat\Psi, \hat H]$.
 We consider the system to be in a classical thermal state of the canonical ensemble with probabilities for a field configuration $\phi$ given by 
 $p[\phi] = \exp(-\beta H[\phi])$. As in the quantum case, we may linearize and diagonalize the theory for sufficiently low classical depletion 
 $\Delta = \sum_{k\neq0} \phi_k^* \phi_k$. 
 The linear transformations to the classical quasiparticles (normal modes) are completely analogous,  preserving the correspondence between the equations of motion 
 of the quantum and classical case.
 This similarity between the quantum and classical linear theory  
 is well known for the common-place approach of linearizing \emph{after} deriving the equations of motion, i.e.~writing either
 $\hat \Psi(x,t) \rightarrow \phi_0(x,t)+ \delta \hat\phi(x,t)$ in the quantum case and $\phi(x,t) \rightarrow \phi_0(x,t)+ \delta \phi(x,t)$
 in the GPE case (for the homogeneous time dependent case considered here, $\phi_0(t)$ is constant in space).
 Thus, we obtain the equations of motion for the classical theory without further calculation by making a replacement 
\begin{equation}
\label{Eq:replace}
\hat A, \hat B \rightarrow A, B
\end{equation} in the corresponding equations from the quantum theory~(\ref{eq:firstordersys}), where the classical quasiparticles 
are defined by a similar linear Bogoliubov transformation of Fourier modes of the field as in the 
quantum case,
\begin{eqnarray} \label{Eq:Bogo}
a   &=& \lambda_k^* A\1 + \gamma_k (B\1)^*	 ,
\end{eqnarray}
where again we have set $\phi_{\mathbf k} = a$, with the coefficients $\gamma_k$ and $\lambda_k$ similarly given by (\ref{Eq:lambda}) and (\ref{Eq:gamma}). 
We identify the mode intensity $\mathbb{E}(I_a)$ with $\langle a^* a\rangle_{cl}$, and similarly for the intensity correlations. 
Within this classical treatment of the BEC, 
 we find again by evaluating a Gaussian integral that $  \braket{A^*A}_{cl}  = n_{k,cl}^{th}$ and $ \braket{A B}_{cl} = 0$ for the initial thermal state. 
However now the energy is equally partitioned over all the field modes,
\begin{equation}
n_{k,cl}^{th} = \frac{k_B T}{\hbar \omega_k} \; 
\end{equation} 
in accordance with the high temperature limit of the Bose-Einstein statistics with vanishing chemical potential. 
The equipartition of energy is indeed observed in equilibrating micro-canonical simulations of the projected GPE~\cite{Davis2001, Davis2002}.
Note that furthermore it has been argued there that the projected 
GPE at finite temperature provides indeed a (classical) theory of the highly occupied modes of an actual Bose-Einstein 
condensate. We do however not rely on a physical interpretation of the classical theory of this sort; 
instead, we just use it as a tool providing a definite classical reference,
to define and isolate the quantum effects.

\begin{figure*}[htp]
  \centering
  \subfigure[]{\includegraphics[scale=0.8]{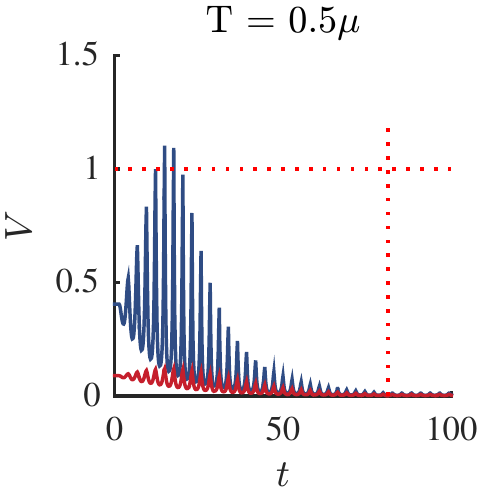}}\quad
  \subfigure[]{\includegraphics[scale=0.8]{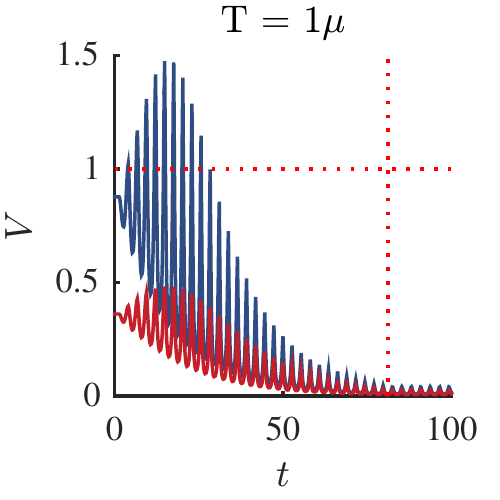}}
  \subfigure[]{\includegraphics[scale=0.8]{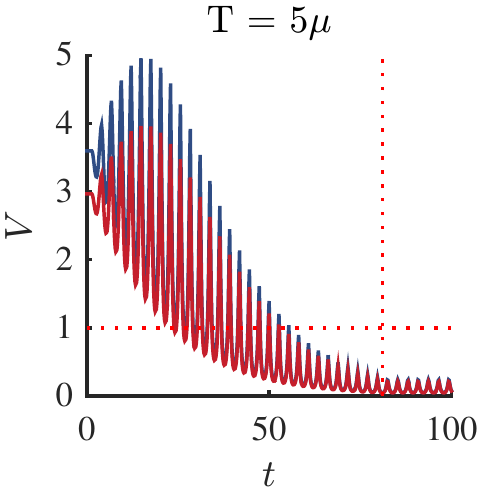}}
  \subfigure[]{\includegraphics[scale=0.8]{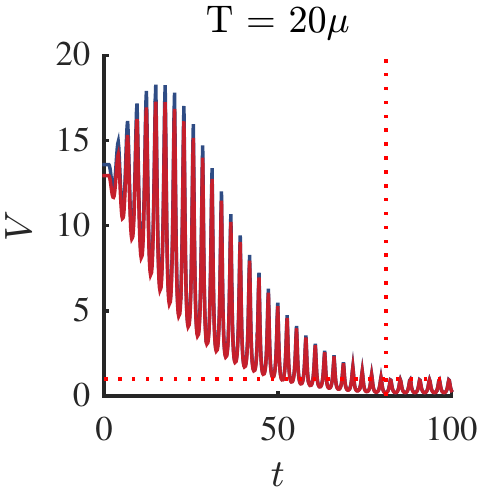}}
  \caption{Time dependence of $V$ in the quantum theory (blue) and classical reference (red) for first resonant mode at (a) $T=0.5\mu$, (b) $T=1\mu$, (c) $T=5\mu$,  and (d) $T=20\mu$. The driving oscillation in both cases is turned off after 40 periods as indicated by the (red) dotted vertical line. The (red) dotted horizontal line indicates $V=1$. For $V<1$ one speaks of sub-Poissonian number squeezing.}\label{fig:Vkrest}
  \label{Vkrest}
\end{figure*}

%%%%%%%%%%%%%%%%%%%%%%%%%%%%%%%%%%%%%%%%%%%%%%%%%%%%%%%%%%%%%%%%%%
\subsection{Correlations} 
%%%%%%%%%%%%%%%%%%%%%%%%%%%%%%%%%%%%%%%%%%%%%%%%%%%%%%%%%%%%%%%%%%
Using the Bogoliubov transformation~(\ref{Eq:Bogo}) we obtain
\begin{eqnarray}
\label{Eq:nkcl} n_{k,cl}^2 &:=&  \left( \braket{a^* a }_{cl}\right)^2 = (2|\gamma_k|^2+1)^2 (n_{k,cl}^{th})^2 ; \\
\label{Eq:mkcl} \left| m_{k,cl}\right|^2 &:=&  \left| \braket{ a b }_{cl}\right|^2   = 4(|\gamma_k|^2+1) |\gamma_k|^2  (n_{k,cl}^{th})^2 .
\end{eqnarray}
Notice the absence of terms leading to the spontaneous generation of field fluctuations within this classical treatment of the excitations. 
It is also worth stressing that even in the absence of a modulation, $\beta_k = 0$, the thermal noise is correlated by the interaction term in the real particle basis, which contributes to the (thermal) depletion.
However, for $T\rightarrow 0$ this depletion goes to zero in contrast to the quantum formulation. From equation~(\ref{Eq:nkcl}) and (\ref{Eq:mkcl}) it is immediate that $\vert m_{k,cl} \vert < n_{k,cl} $. Hence the \emph{mode CSI cannot be violated} in this classical theory. This is in agreement with the mathematical Cauchy-Schwarz theorem.

The Wick theorem for the evaluation of higher correlation functions now follows from Isserlis' theorem~\cite{Isserlis:1918aa} directly. The fourth moments needed in expression~(\ref{Eq:TMV_fun}) for the two-mode variance 
are  $ \mathbb E (I_a I_a) =\mathbb E (a^* a a^* a) = 2\braket{a^*a}_{cl}^2$ and $  \mathbb E (I_a I_b) = \braket{a^*a b^*b}_{cl} 
= \braket{a^*a}_{cl}^2+\left|\braket{ a b}_{cl} \right|^2$. (Note, comparing the classical with the quantum theory no corrections due to normal ordering are needed in order to apply Wick's theorem.)

%%%%%%%%%%%%%%%%%%%%%%%%%%%%%%%%%%%%%%%%%%%%%%%%%%%%%%%%%%%%%%%%%%%
%
\section{Comparison between classical and quantum theory for specific observables
 \label{Sec:Comparison}}
%
%%%%%%%%%%%%%%%%%%%%%%%%%%%%%%%%%%%%%%%%%%%%%%%%%%%%%%%%%%%%%%%%%%%
Finally we apply our findings to recent experimental procedures to detect sub-Poissonian statistics, intensity CSI and mode CSI.  We demonstrate that \emph{none} of these proposed experimental observables can establish the nonclassciality of the excitations in a BEC.
%%%%%%%%%%%%%%%%%%%%%%%%%%%%%%%%%%%%%%%%%%%%%%%%%%%%%%%%%%%%%%%%%%%

\subsection{Sub-Poissonian statistics \label{subSEC:subP}}
%%%%%%%%%%%%%%%%%%%%%%%%%%%%%%%%%%%%%%%%%%%%%%%%%%%%%%%%%%%%%%%%%%%
Independent of the underlying theory the TMV, see equation~(\ref{Eq:TMV_fun}), can be extracted experimentally via TOF and\,/\,or in-situ measurements. From a single experiment one can extract $I_aI_a$, $I_aI_b$ and $I_a$. By repeating the experiment and then averaging over the extracted quantities one can then get the corresponding expectation values. As pointed out above if the TMV is smaller than $1$ the statistics of the process is referred to as sub-Poissonian.

Our classical reference predicts 
  \begin{eqnarray}
  \label{Eq:Vclass}
  V_{cl}= \frac{n_{k,cl}^{th}}{2 \vert \gamma_k \vert^2 +1} \;.
  \end{eqnarray} 
  Note that $V$ and $V_{cl}$ are mathematically inequivalent expressions, but physically refer to the \emph{same} observable. 
Thus we obtain sub-Poissonian statistics for
\begin{equation}
|\gamma_k^2(t)| > n_{k,cl}^{th}(T/2) - \frac{1}{2} \label{Eq:subPoissonianClassical},
\end{equation} 
within a completely classical treatment of small fluctuations in a BEC. 

Comparing the quantum result (\ref{Eq:entangled2}) with the classical result (\ref{Eq:subPoissonianClassical}), it appears that it is possible in both cases to produce sub-Poissonian statistics for a sufficiently low temperature and suitable modulation. 
We compare the time dependent value of $V$ for a resonant mode undergoing periodic modulation for both the quantum and classical 
cases in Fig.~\ref{fig:Vkrest}. Panel (a) shows the case of $T=0.5\mu$ and panel (b) $T=2\mu$. 
There are three important observations that can be made from these plots, which are true for both theories. 
First, $V$ is in general time-dependent due to the form of $\gamma^2$ given in Eq.~(\ref{Eq:gammasq}). 
Second, for sufficiently long driving the oscillations become suppressed and $V$ approaches zero. 
The oscillations in $V$ for the quantum case also suggest that the entanglement citerion (\ref{Eq:entangled2}) is time-dependent. 
It should be noted that this refers to mode entanglement in contrast to particle entanglement. 
Third, the classical value of $V$ is always below the quantum value. This follows 
from comparing the criteria~(\ref{Eq:subPoissonianClassical}) 
and~(\ref{Eq:entangled2}) and given that $n_{k,cl}^{th} < n_k^{th}$ at finite temperature. 
The physical explanation is that our classical reference is a wave theory and as such is 
missing the shot noise due to the discrete excitations (in this case atom numbers), i.e.~the first term on the right hand side of 
$Var(\hat n_k) =\braket{\hat n_k}+ \braket{:\hat n_k \hat n_k:} - \braket{\hat n_k}^2 = n_k + n_k^2$. The classical reference only contains the final term $n_k^2$ that has been associated by Einstein to the wave character of the atoms~\cite{Einstein1925}. 
 The shot noise term is necessary for a (super-)Poissonian two-mode variance for low occupation numbers. 
To see this, we make use of an alternative form of expression~(\ref{Eq:TMV_fun}), $V = (Var(I_a) - Cov(I_a,I_b))/\mathbb E(I_a)$ 
to bound $V$ from above by  $V \le Var(I_a)/\mathbb E(I_a)$ for positive or vanishing correlations. 
Since classically the variance scales with the square of the occupation number, $V$ will always be sub-Poissonian for low enough occupation numbers. 
Indeed, for equilibrium and vanishing interactions, (\ref{Eq:Vclass}) becomes $V_{cl}=n_{k,cl}^{th}$ showing the threshold is an occupation of unity. 

We would like to stress that the absence of the shot noise in the classical theory is \emph{not} always the cause of the obtained sub-Poissonian statistics. At high initial temperature $T\gg1$ the shot noise becomes insignificant and the classical and quantum results approach each other. For strong enough driving, sub-Poissonian statistics is possible in both theories, even if the initial thermal state exhibits super-Poissonian fluctuations.

We conclude that a TMV smaller than 1, and\,/\,or sub-Poissonian statistics is \emph{not} a sufficient nonclassicality criterion to rule out our specific classical reference.

%%%%%%%%%%%%%%%%%%%%%%%%%%%%%%%%%%%%%%%%%%%%%%%%%%%%%%%%%%%%%%%%%%%
\subsection{Violation of intensity Cauchy-Schwarz inequality for time-resolved TOF measurements~\label{subSEC:intCSI}}
%%%%%%%%%%%%%%%%%%%%%%%%%%%%%%%%%%%%%%%%%%%%%%%%%%%%%%%%%%%%%%%%%%%
As pointed out in (Sec.~\ref{Sec:NonClassicalityCriteria}) the so-called violation of the intensity CSI, defined as auto-correlations exceeding cross-correlations, is only possible if the auto-correlations are normal ordered in the quantum theory or equivalently $I_a(I_a-1)$ is measured instead of $I_a^2$ for the auto-correlations. If $V<1$ a comparison of the cross-correlations with the auto-correlations in the reduced form $I_a(I_a-1)$ appears in both the quantum and classical case. A distinction between the quantum theory and the classical reference as in coincidence counting experiments is then not available.

A time-resolved TOF method as proposed in~\cite{Kheruntsyan2012} is not a suitable detection scheme for this purpose. In the TOF method the atoms are released from the trap, hence the BEC is destroyed, and the wave description for the excitations breaks down.  Collective excitations have been transferred to a finite number of atoms and hence independent of the nature of the collective excitations the measurement of the auto-correlations is destructive: the probability to subsequently measure two excitations with momentum $k$ is proportional to $n_k(n_k-1)$. In this sense the resulting violation of the CSI within a time-resolved TOF measurement of the density-density correlations  is \emph{not} a priori incompatible with a suitable classical description. It is instead equivalent  to $V<1$ and our discussion of the corresponding interpretation of this case applies directly to the results in~\cite{Kheruntsyan2012}.

We conclude a time-resolved TOF measurement as suggested in~\cite{Kheruntsyan2012} is \emph{not} a sufficient nonclassicality criteria for the excitations in a BEC.

%%%%%%%%%%%%%%%%%%%%%%%%%%%%%%%%%%%%%%%%%%%%%%%%%%%%%%%%%%%%%%%%%%%
\subsection{Indirect measurement of violation of mode Cauchy-Schwarz inequality \label{subSEC:modeCSI}}
%%%%%%%%%%%%%%%%%%%%%%%%%%%%%%%%%%%%%%%%%%%%%%%%%%%%%%%%%%%%%%%%%%%

The mode CSI could provide a way to establish the nonclassicality of the system. However, the anomalous density needs to be measured for this purpose. In~\cite{Steinhauer:2015ab,Steinhauer:2014aa} \textit{in situ} density measurements were related to nonseparability criteria for the case of Hawking radiation which in the most simple case also reduce to a mode CSI between upstream and downstream modes~\cite{Busch:2014ab}. A similar analysis was also given in~\cite{Nova:2015aa}.
The basic observables are the atom number densities at different positions in real space $\rho(\mathbf r)$ measured for all $\mathbf r$ by an \textit{in situ} imaging of the condensate. 

The expectation value $\braket{\hat \rho_{\mathbf k} \hat \rho_{-\mathbf k}}$, where $\hat\rho_{\mathbf k'} = \int d\mathbf r \hat\rho(\mathbf r) e^{-i\mathbf k' \mathbf r}$ and $\hat\rho(\mathbf{r})=\hat\Psi(\mathbf{r})^\dag \hat\Psi(\mathbf{r})$, can be obtained experimentally from the ensemble average $\mathbb{E}(\rho_{\mathbf k} \rho_{-\mathbf k}) $ of the Fourier transform $\rho_{\mathbf k'} = \int d\mathbf r \rho(\mathbf r) e^{-i\mathbf k' \mathbf r}$ of the measured density $\rho(\mathbf r)$ over different realisations (shots) of an identical experiment. Note that $\hat \rho_{\mathbf k}$ is not the number operator for mode $\mathbf k$ but a convolution $\hat \rho_{\mathbf k} = \sum_{\mathbf p}  \hat a^\dag_{\mathbf p}  \hat a_{\mathbf k + \mathbf p} $ involving the (dimensionless) real particle annihilation and creation operators $\hat a_{\mathbf k} \equiv V^{-1/2} \int d\mathbf r \hat \Psi(\mathbf r) e^{-i\mathbf k \mathbf r} $, as can be easily verified by using	 $\hat \Psi(\mathbf r) =  V^{-1/2} \sum_{\mathbf k} \hat a_{\mathbf k}  e^{i \mathbf k \mathbf r} $.

Making use of the Bogoliubov approximation~(\ref{Eq:Bapprox}) this is 
\begin{equation} 
\hat \rho_{\mathbf k} \simeq  \sqrt{N} ( \hat a_{\mathbf k} +  \hat a^\dag_{- \mathbf k} ) .
\end{equation}
Here we have assumed that the terms quadratic in the operators are 
 bound by the total depletion.
 We can see immediately that $\hat \rho_{-\mathbf k} = \hat \rho_{\mathbf k}^\dag$ and  $[\rho_{\mathbf k}, \rho_{\mathbf k}^\dag] = 0$. Thus $\braket{\hat \rho_{\mathbf k} \hat \rho_{-\mathbf k}} = \braket{\hat \rho^\dag_{\mathbf k} \hat \rho_{\mathbf k}}$  is a positive quantity and is approximately given by 
\begin{equation}
\braket{\hat \rho_{\mathbf k} \hat \rho_{-\mathbf k}} \simeq N\left[ (\hat a_{-\mathbf k} \hat a_{\mathbf k}+  h.c.) + (\hat a^\dag_{\mathbf k} \hat a_{\mathbf k} +\hat a_{- \mathbf k} \hat a^\dag_{-\mathbf k}) \right] .
\end{equation}

In our previous notation and exploiting the symmetry of the state this reads $ N(\hat a \hat b +\hat a^\dag \hat b^\dag + \hat a^\dag \hat a + \hat a \hat a^\dag)$. 
In the quantum theory, the expectation value is thus 
\begin{equation}
\braket{\hat \rho_{\mathbf k} \hat \rho_{-\mathbf k}} =  N (2\Re{(m_k)} + 2 n_k +1). 
\end{equation}

The behaviour of $m_k = \gamma_k \lambda_k^*(1+2n_k^{th})$ is determined by the complex factor
\begin{eqnarray} \label{Eq:needed}
\gamma_k \lambda_k^* &=&  
   - |u_k v_k| (2|\beta|^2 + 1)  \\ 
  \nonumber
   &+& \sqrt {|\beta|^2+1} |\beta|\left[2 v_k^2 \cos(2\omega t +\delta_k) + e^{-i 2\omega t - i\delta_k}\right].
\end{eqnarray}
(Note that $v_k < 0$ was used in the first term.)
At times $t_m$ such that $(2\omega t_m +\delta_k)\mod 2\pi = \pi$, this reduces to
\begin{equation} \label{Eq:needed}
\gamma_k \lambda_k^* = - |u_k v_k| (2|\beta|^2 + 1) - \sqrt {|\beta|^2+1} |\beta|(2 v_k^2  + 1), \quad
\end{equation}
a negative real number. Since all terms enter the rhs of equation~(\ref{Eq:needed}) appear with the same sign at $t_m$ the term  $|\gamma_k \lambda_k^*|$ is largest at this time.
 Thus, when measured at this particular time $t_m$, we can replace the real part by a minus sign,
 \begin{equation} \label{Eq:JeffQT}
\braket{\hat \rho_{\mathbf k} \hat \rho_{-\mathbf k}} = 2 N ( n_k - |m_k|) + N. 
\end{equation}
We can then see that -- assuming the quantum theory -- at $t_m$ the mode CSI violation~(\ref{Eq:modeCSI}) is equivalent to 
\begin{equation}
 \braket{\hat \rho_{\mathbf k} \hat \rho_{-\mathbf k}} < N . 
 \end{equation}
Keeping track of higher order terms shows that \emph{one} condition for this analysis to hold is $\Delta \ll N/n_k$, which has to be compared to $\Delta \ll N$ for the validity of standard Bogoliubov theory, where $\Delta$ is the depletion of the condensate~(\ref{Eq:depletion}).
 
Note that $ \braket{\hat \rho_{\mathbf k} \hat \rho_{-\mathbf k}} = N $ in the limit of zero temperature, no driving and vanishing interactions due to the nonvanishing commutator. (The quantum noise causes the measured in-situ density to fluctuate.) Thus, the observable consequence of entanglement in this experiment is a measured  \emph{supression} of these fluctuations. The absence of quantum noise in any classical reference already indicates that such a threshold does not exist in our classical reference.

Indeed, we get by calculation that
\begin{equation}
 \mathbb{E}(\rho_{\mathbf{k}}\rho_{\mathbf{-k}})  = \braket{\rho_{\mathbf k}  \rho_{-\mathbf k}}_{cl} = 2 N ( n_k - |m_k|),
\end{equation}
hence it is also possible to obtain $\mathbb{E} ( \rho_{\mathbf k}\rho_\mathbf{-k}  ) < N $. Very much like in the TMV case there are commutator terms only appearing within the quantum theory that make the observation imply a violation of the mode CSI.

In summary,  an indirect measurement of the mode CSI of the type suggested in~\cite{Steinhauer:2015ab,Steinhauer:2014aa} is \emph{not} a sufficient non-classicality measure when applied to the case of excitations in a homogenous parametrically excited BEC. \\

%%%%%%%%%%%%%%%%%%%%%%%%%%%%%%%%%%%%%%%%%%%%%%%%%%%%%%%%%%%%%%%%%%
%
\section{Conclusions\label{Sec:Conclusions}}
%
%%%%%%%%%%%%%%%%%%%%%%%%%%%%%%%%%%%%%%%%%%%%%%%%%%%%%%%%%%%%%%%%%%
In this paper we have taken the viewpoint that nonclassical effects in the BEC are those that are 
incompatible with the results obtained using an ensemble of classical trajectories given by solutions of the GPE. 
For excitations in a parametrically excited BEC we showed that \emph{observable} strong number correlations as indicated by $V<1$ 
(related to P-nonclassicality and even nonseparability of atomic modes in the quantum theory) and an indirect measurement of the intensity CSI 
 as suggested in~\cite{Steinhauer:2015ab,Steinhauer:2014aa} are compatible with this classical theory. Similarly, we argued that 
 a violation of the intensity CSI in a time-resolved TOF measurement is compatible with the classical picture of an atomic cloud 
 after destruction of the BEC.  
 Nevertheless, in the context that the Bogoliubov theory is a valid approximation, a measurement of $V<1$ 
 (or equivalently a violation of the intensity CSI) is sufficient for P-nonclassicality and nonseparability, 
 both of which are criteria that assess a given quantum state. 
 However, these criteria are not suitable to theoretically single out the spontaneous process of amplified 
 vacuum noise for analogue gravity studies. 

We conclude that the direct experimental verification of the quantumness of the fluctuations remains an open challenge for future BEC experiments in general. For this purpose we propose that additional observables, including noncommuting ones, e.g.~density \emph{and} phase fluctuations, have to be measured. 
Returning to the particular setup discussed above, a parametrically excited condensate, we would like to point out that such a system can be used to mimic models of cosmological particle production in table-top experiments~\cite{Jain2007}. As argued in this paper, even within  highly controllable and repeatable BEC experiments one is for now facing a similar dilemma to the one of establishing the quantum origin of the fluctuations seeding our universe~\cite{Maldacena:2015aa}. So far in both cases a suitable Bell-type experiment, that would for once and all resolve this issue, is absent. 

\acknowledgements
We acknowledge Antonin Coutant, Ioannis Kogias, Peter Kruger, Bill Unruh, Madalin Guta, and Joerg Schmiedmayer for fruitful discussions, and Antonin Coutant, Dieter Jaksch and Matt Davis for feedback on the manuscript.  A.F. would like to further thank Sammy Ragy, Katarzyna Macieszczak, Peter Gr\"unwald and Matteo Marcuzzi for enlightening discussions. S.W. acknowledges financial support provided under the Royal Society University Research Fellow, the Nottingham Advanced Research Fellow and the Royal Society Project grants.

%\newpage
%\bibliographystyle{apsrev}
%\bibliography{references}
\end{document}